# Sub-monolayer structures of Ag overlayers on Ge(111): experimental observations and first-principles study

Shree Ram Acharya[1], Cory H. Mullet[2,*], Jason A. Giacomo[2,†], Duy Le[1,3], Shirley Chiang[2], and Talat S. Rahman[1,3,‡]

[1]Department of Physics, University of Central Florida, Orlando, FL 32816-0115, USA

[2]Department of Physics, University of California, Davis, CA 95616-8677, USA

[3]Renewable Energy and Chemical Transformations Cluster, University of Central Florida, Orlando, FL 32816-0115, USA

We present a joint experimental and theoretical determination of structures of Ag adatoms on the Ge(111) surface using low energy electron diffraction, low energy electron microscopy, scanning tunneling microscopy, and density functional theory-based calculations, as functions of coverages and temperature. Experimentally for clean Ge(111), c(2×8) and (2×1) phases occur, while Ag overlayers cause (4×4), ($\sqrt{3}\times\sqrt{3}$)R30º and (3×1) surface structural phases. The dependence of the growth behavior of these different phases was examined as a function of temperature, Ag deposition rate and coverage, substrate step density, and history of temperature cycling. First-principles calculations of the electronic and geometric structures and vibrational dynamics show the Ge(111)-c(2×8) configuration with Ge adatoms adsorbed on three-fold hollow ($T_4$) sites to be the energetically most favored phase of the Ge(111) surface, among unreconstructed Ge(111), reconstructed Ge(111)-2×1, and Ge(111)-c(2×8) structures. The Ge(111)-Ag(3×1) overlayer of the system has Ge atoms forming a honeycomb chain on a missing top layer reconstructed surface, with metal at $\frac{1}{3}$ ML coverage in channel. The Ge (111)-Ag($\sqrt{3} \times \sqrt{3}$)$R30^0$ overlayer contains one monolayer Ag forming inequivalent Ag triangles in a surface unit cell on the missing top layer reconstructed Ge(111) surface. The Ge(111)-Ag(4× 4) overlayer formed at low Ag coverage contains two triangular subunits at different heights: one with six Ag adatoms and the other with three Ge adatoms on the intact double layer Ge(111) surface. The temperature and coverage dependent surface phase diagram, obtained by minimizing the surface free energy, captures the main features of the experimental phase diagram.

---

[*] Current address: Intel Corporation, Hillsboro, OR 97124.

[†] Current address: Air Quality Research Center, University of California, Davis, CA 95616-8677.
[‡] Corresponding author email address: Talat@ucf.edu (Talat S. Rahman)

## I. INTRODUCTION

A system with sub monolayer Ag deposited on the most compact (111) surface of Germanium [Ag/Ge(111)] has gained interest due to the formation of various one- and two-dimensional structures of the system by manipulating temperature and metal coverage (see [1] and references therein). In fundamental aspects, such structures are relatively convenient to analyze to get an understanding about interactions of Ge and Ag atoms, to find atomic configurations at which they are stable as a function of coverage and temperature, and to explore the source of phase transformations from one structure to another. Such a basic experimental and theoretical understanding can pave the way for designing complex models of the growth processes of metallic thin films over semiconducting surfaces.

We first present our experimental observations of the structural phases of the Ag/Ge(111) system, as measured primarily by low energy electron diffraction (LEED) and low energy electron microscopy (LEEM). Similar to earlier published results [1], we observe several structural phases, such as $(\sqrt{3} \times \sqrt{3}) R30^0$, (4× 4), and 3×1, as a function of temperature and coverage. Below temperatures of 300°C, the clean Ge(111) surface displays the c(2×8) reconstruction, while between $300^0$C and $420^0$C, a (2×1) LEED pattern appears. As the Ag coverage is increased, the (4×4) and (3×1) phases form up to coverages of 0.375 ML. We show in this paper details of the formations of regions with these different phases, particularly how these structural phases nucleate at steps, on terraces, and at defects on the surface. At higher coverage, from 0.375 to 1 ML, the (4×4) and $\sqrt{3}$ phases occur for temperatures <$500^0$C. The (3×1) phase also appears in narrow temperature ranges from $420^0$C to $570^0$C, and depending on coverage, can appear with the (4×4) or $\sqrt{3}$ phases. Above $570^0$C, we observe only a (1×1) LEED pattern, indicating no specific ordered structure.

The construction of atomic models of observed structures is an important first step to further explore properties of a system. However, there is no consensus on the geometrical models of observed overlayers of the Ag/Ge(111) system. For $\sqrt{3}$ overlayer, the honeycomb- chained- trimer (HCT) model corresponding to the honeycomb pattern with two bright spots per $\sqrt{3}$ unit cell [2-4] and inequivalent-triangle (IET) configurations [5] of hexagonal pattern with two types of Ag triangles of different sizes in $\sqrt{3}$ unit cell that correspond to the bright and dark spots in the image from scanning tunneling microscopy (STM) are proposed. The possibility of obtaining one or the other periodic pattern by manipulating the tunneling bias in STM is reported in ref. [6]. For (4× 4), based on STM data and charge transfer insight, Hammar et al. [2] proposed a model with Ag and Ge adatoms on double layer beneath on which Ag at 6/16 monolayer (ML) coverage adsorbs on top of Ge on one half of unit cell and Ge on the other half occupy $T_4$ sites with rest atom between them. A slightly modified model is proposed by Spence et al. [7] based on detailed STM measurement in which Ge adatoms form trimer. Collazo-Davilla et al. [8] proposed a missing top layer reconstructed model with six Ag atoms on Ge substitutional sites and ring-like assembly of nine Ge atoms on two triangular subunits based on direct method to surface X-ray diffraction data. Based on the observation of missing top layer surface due to laveling of 4×4 and $\sqrt{3}$ tetraces and symmetry in images on STM measurement, Weitering et al. [9] proposed a model with (2×2) sublattices of Ge trimers and Ag atoms centered at $T_4$ sites and $H_3$ sites, respectively and second-layer Ge restatoms. For Ag-(3×1)/Ge(111), various models can be candidates from the Ag-(3×1)/Si(111) surface: Seiwartz-chain model [10-12] which is a missing top layer model that has five-fold chains along [1 $\bar{1}$ 0] direction with metal atoms on channel, the extended Pandey model [12,13] formed by extension of (2×1) $\pi$-bonded chain model and honeycomb channel (HCC) model [8,14-17] in which surface atoms form a honeycomb chain. The former two models could explain the metal insensitive LEED I-V curve [18] but can not expain surface band structure and STM images [19,20]. The HCC model is shown computationally to be the energetically favored with double bond formation and reproduces ARPES and STM results [16,17].

To get an atomic model of observed ordered surface structures, their electronic structure and thermodynamic stability in a single computational setup so that energetics are comparable, the density functional theory [21,22] based calculation is performed in this study. The bonding characteristics are explained by calculating the electronic states and charge density distributions. To compare energetics of overlayers with different number of atoms, the surface free energy which quantifies the amount of energy required to form a surface, either by breaking bonds of bulk stacking or adding adatoms is calculated. In addition to the general practice of such calculations at 0 K, the contribution of vibrational entropy of atomic motion is incorporated in this study by coupling the static DFT with vibrational calculations. The vibrational calculation of the energetically favored configuration of each overlayer provides the thermodynamic stability of the model and the temperature dependent surface free energy. For a given temperature and coverage, the surface free energies obtained for a combination of different phases are minimized to determine the favored phase in the given temperature and coverage leading to construct the surface phase diagram.

In the rest of this paper, we present first the details of the experimental set up in section 2, the experimental results in section 3, computational set up in section 4, computational results and discussion in section 5, and conclusions in section 6.

## II. EXPERIMENTAL DETAILS

Measurements were carried out in a ultrahigh vacuum (UHV) system consisting of three connected chambers housing several commercial instruments, including a LEEM (Elmitec GmbH), STM (Oxford Instruments), and x-ray photoemission spectrometer (Vacuum Generators) [23].

Ge(111) samples were cut from Sb-doped Ge(111) 2-inch wafers purchased from MTI Crystal, with resistivity ~0.25 $\Omega$-cm and polished on one side to within 0.5° of the (111) surface. Ge(111) crystals were cleaned in ultrahigh vacuum with repeated cycles of $Ar^+$ bombardment (250 eV, 5 µA) of the unheated sample for 15 minutes, followed by annealing the sample between 800°C and 830°C for 10 minutes (30 minutes for the last anneal before imaging in LEEM or STM). The energy of the $Ar^+$ beam was 0.25 keV. Sputtering and annealing cycles were performed in the analysis chamber (base pressure $2\times10^{-10}$ Torr) before transferring the sample to the LEEM chamber (base pressure $1\times10^{-10}$ Torr) or STM chamber (base pressure $4\times10^{-10}$ Torr). Samples were sputtered and annealed until a clean c(2×8) LEED pattern was obtained, as shown in FIG. 1 (a). For reference, the LEED pattern of the Ge(111)-(2×1) reconstruction is also shown in FIG. 1 ((b)-(c)). The clean Ge(111)-c(2×8) reconstruction transforms to a (2×1) reconstruction at 300°C, with some hysteresis [24]. The (2×1) LEED pattern has characteristic oblong features at (½, ½) (FIG. 1 ((b)-(c)), pointed out with blue arrows, the shape and intensity of which vary with temperature [24]. X-ray photoemission spectroscopy (XPS) was also occasionally used to verify that the sample was free from contaminants.

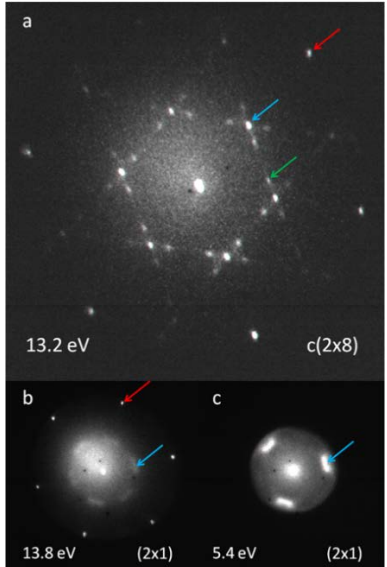

FIG. 1. LEED pictures of clean, reconstructed Ge(111) surface. Red arrows point to first-order spots, blue arrows to half-order spots, and green arrows to eighth-order spots. (a) Ge(111)-c(2×8) near room temperature. (b)-(c) Ge(111)-(2×1) at 370°C.

Sample heating was monitored through a K-type thermocouple with the junction pressed against the back of the sample. The thermocouple was then calibrated with an infrared pyrometer with the emissivity set to 0.42. The emissivity was found by calibrating the pyrometer to the melting point of an old Ge sample. The thermocouple calibration was then checked by measuring the temperature for the c(2×8) to (2×1) phase transition at 300°C on clean Ge(111) [24].

Ag was deposited via direct evaporation from an Ag wrapped tungsten filament which was resistively heated. The deposition rate was calibrated in the LEEM with a Ge(111) sample by measuring the time to complete the (4×4) phase (at 0.375 ML), $\sqrt{3}$ phase (at 1.00 ML), or by the partial completion of one or both of these phases, as shown in FIG. 2. A range of deposition rates (0.005-1.5 ML/min) was employed depending upon the phenomena under study.

### III. EXPERIMENTAL OBSERVATIONS

The growth of Ag on the reconstructed Ge(111)-c(2×8) surface was studied by dosing Ag in the LEEM while imaging the surface at various temperatures. No evidence of Ag dosing could be seen in the LEEM images before about 0.1 ML of coverage. After 0.1 ML of Ag coverage, a darkening and broadening of the steps on the surface can be seen in the LEEM images. We examined the different growth behaviors of the various structural phases [(4×4), $\sqrt{3}$, and (3×1)] of Ag/Ge(111) as a function of the substrate temperature during the growth process.

For growth of Ag on Ge(111) at temperatures below ~185$^0$C, the (4×4) phase does not form. Instead the $\sqrt{3}$ phase appears to grow as dark features at low coverage until it covers the surface at ~0.5 ML (not shown). Evidently, reduced mobility of the atoms at lower temperature causes the growth of small domains and clusters on the surface. However, the $\sqrt{3}$ phase must be filled with vacancies, as it will continue to accept more Ag up to a total coverage of 1.0 ML before the second layer and 3D islands begin to form. There is the possibility of a second, lower density, $\sqrt{3}$ structure, similar to the behavior of Pb/Ge(111), which has two different $\sqrt{3}$-structures, the dilute α and dense β phases [25,26]. No such indication has been found in the literature, however, and the LEEM data do not show any contrast changes that might indicate two separate structures.

Dosing Ag onto the Ge(111) at 200°C allows us to look at the region of the phase diagram where the reconstructed c(2×8) substrate phase and the (4×4) and $\sqrt{3}$ Ag phases all coexist, up to a coverage of ~0.4 ML (FIG. 2). At this temperature, the mobility of adatoms, especially Ge, on the surface is limited. This in turn reduces the faceting of the Ge surface and creates a barrier to the formation of the (4×4) phase. It then takes a sufficient build-up of Ag adatoms on the surface to induce a phase transition which then produces the higher density $\sqrt{3}$ phase. At 200°C the (4×4) phase forms at the edges of the $\sqrt{3}$ phase as seen in Fig. 3, indicating that there is still some Ge mobility at the phase boundaries.

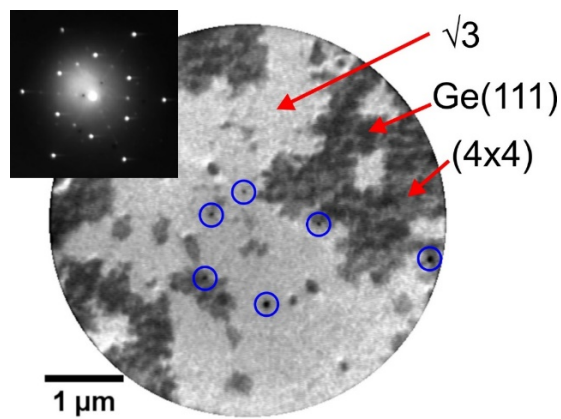

FIG. 2. Ag dosed at 200°C does not completely cover the surface, and the two Ag phases, $\sqrt{3}$ and (4×4), coexist up to ~0.4 ML. This image shows the coexistence of the two phases and the Ge substrate at Ag coverage of 0.33 ML. The blue circles surround six dark spots that are due to dead spots in the microchannel plate; these are present in all LEEM images shown in this paper. The inset shows the LEED pattern displaying both the $\sqrt{3}$ and the (4×4) spots.

FIG. 3 (a-d) shows a progression of LEEM images as Ag was dosed up to 0.33 ML at $250^0$C. LEED was done at various points during dosing to confirm that this is the (4×4) phase. Ag begins forming the (4×4) phase along the steps of the substrate. As the coverage increases, the images show the (4×4) phase growing out from the steps in both directions. The substrate steps remain in their original positions as the (4×4) phase spreads out onto the terrace from both the upper and lower step edges. This is not step flow growth but rather a nucleation of the phase at the steps, instead of forming clusters on the terraces. The growth continues until the surface is completely covered by the (4×4) structure.

Another interesting observation on the dynamics of sub-monolayer coverages of Ag on Ge(111) has to do with the diffusion of Ag atoms on the surface. Two groups have measured the diffusivity of Ag on the clean, reconstructed Ge(111) surface by watching the evolution of Ag islands deposited through masks. Suliga and Henzler found, using scanning Auger electron spectroscopy (AES), that the Ag diffusion coefficient along the steps was two orders of magnitude larger than that on a smooth Ge(111) face [27]. Similarly, Metcalfe and Venables used scanning AES to study the effect of diffusion on the surface [28]. They also incorporated biased secondary electron imaging (b-SEI). In their b-SEI images the Ag island shapes led to the conclusion that Ag was diffusing more rapidly along steps on the surface. They also compared their results to previous Ag/Si(111) studies and note that Ge adatom movement (as deduced from the observed faceting [27,29]) could explain the differences in the two systems and the increased Ag diffusion along steps in the Ge(111) surface. Ge adatom mobility (or the lack of it) at lower temperatures as indicated by the lack of faceting could also explain the different growth characteristics observed in LEEM at these temperatures, as discussed above.

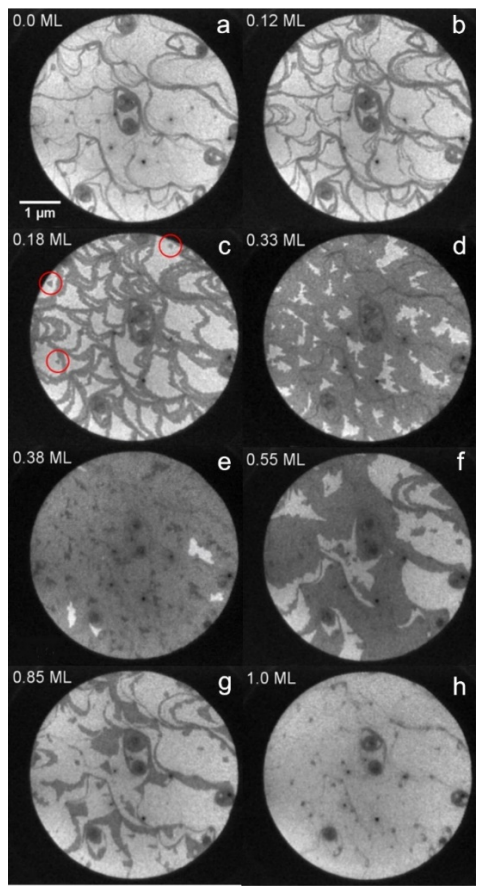

FIG. 3. Ag is dosed onto Ge(111) at 250°C. (a)-(d) The dark (4×4) phase is seen to form first at steps and then to grow outward onto the terraces. At 0.18 ML a few isolated islands of (4×4) can be seen growing (red circles). The (4×4) phase completes at Ag coverage of 0.37 ML. (e)-(h) After the completion of the dark (4×4) phase, the light gray $\sqrt{3}$ phase begins to grow. Notice that in contrast to the way the (4×4) phase grows on the surface, the $\sqrt{3}$ phase nucleates on a terrace and then grows to fill the terrace, with the layer completing at 1.0 ML.

The lower barrier to diffusion at the steps seems counterintuitive given that the phase is expected to begin growing where adatoms slow down enough to nucleate into ordered structures. From this point of view the observation of the phase growth at the steps would suggest that the steps are acting as sinks for adatoms, with higher mobility on the terraces. However, this is a simplistic view of the system and the structure of the phase itself needs to be considered. The (4×4) structure incorporates both Ge and Ag atoms on an intact double layer [2]. Thus, efficient growth of the (4×4) phase requires mobility of both Ag and Ge adatoms. The faceting observed by Suliga and Henzler [29] indicates Ge adatoms are mobile at the steps. Thus, the phase begins to form at the step facets where both atomic species are mobile.

As dosing of Ag continues, the (4×4) phase completes and a denser $\sqrt{3}$ phase begins to form. There has been some controversy over the saturation coverage of the $\sqrt{3}$ phase and hence the structure of the phase. Two early studies found the $\sqrt{3}$ coverage using AES break points as 0.85 ML [30] and 0.82 ML [27]. A later 2-D X-ray structure analysis proposed a model with a saturation coverage of 0.72 ML [31]. However, after STM imaging was successfully applied to the Ag/Si(111) system, [32] it was used to study Ag/Ge(111) and strong similarities were recognized between the two systems [2]. The honeycomb chained trimer (HCT) model had already been proposed for the Ag/Si(111) system which had a coverage of 1.0 ML [33]. Two more studies utilizing core-level photoelectron spectroscopy [34] and AES [28] also concluded Ag/Ge(111)-$\sqrt{3}$ had a HCT structure. A STM study by Spence and Tear, [7] which showed well separated (±1.0 V) occupied and unoccupied images that compared perfectly with similar STM images on Ag/Si(111), [32] solidified the HCT model as the accepted structure. The STM study by L-W Chou et al. [5] has shown an inequivalent triangle (IET) structure with two types of Ag triangles having different sizes in the surface unit cell and forming hexagonal periodicity on the surface. Another STM study by H. M. Sohail et al. [6] observed HCT and IET structures based on tunneling bias. LEEM's strength lies in observing the contrast differences caused by ordered patterns on the surface but not in determining the structures that produce them. Thus, LEEM is not suited to determining surface structures and cannot definitively say which saturation coverage is correct, but it seems likely that the true coverage is 1.0 ML, corresponding to the HCT model.

FIG. 3 ((e)-(h)) shows the progression of the $\sqrt{3}$ phase as it begins to form at 0.37 ML until it completely covers the surface. Immediately we notice that the growth is different from the (4×4) phase. The $\sqrt{3}$ phase depends less on the step structure and more freely grows out onto the terraces. As for the (4×4) data, this observation is backed up by diffusivity measurements done by Suliga and Henzler and also Metcalfe and Venables [27,28]. Both of these groups, through different techniques, have found that Ag adatoms have higher mobility on the Ag-(4×4) surface than on the Ge(111)-c(2×8) surface with no preferred mobility in the step direction. Also, Ge atoms have already been incorporated into the (4×4) structure and the increased Ag density then provokes the transition to the higher coverage $\sqrt{3}$ phase. This is why the $\sqrt{3}$ phase grows as larger islands on the surface rather than out from the steps as the (4×4) does. However, there appears to be some reduction in the diffusion of Ag adatoms across steps (presumably due to an Ehrlich-

Schwoebel barrier) as the $\sqrt{3}$ domains are typically bound by steps. A more complete picture of these phenomena is formed when we look at other dosing temperatures and different samples.

Growth of the Ag (4×4) phase at 410°C is shown in FIG. 4. Initial growth of Ag (4×4) first appears in LEEM images at step edges for < 0.1 ML coverage (FIG. 4 (a)). Growth proceeds outward from both edges of the step. Nucleation on terraces is also observed (FIG. 4 ((c)-(d)). The lag time between nucleation at steps and nucleation on terraces is greater at higher substrate temperatures during deposition (FIG. 4) than at lower temperatures. For growth both at terraces and at steps, the (4×4) domain boundaries are characterized by distinct faceting. The substrate temperature during deposition also affects the number density of Ag islands on terraces, which is higher for deposition at lower temperature. In addition, at the same temperature, a higher deposition rate yields a higher number density.

In addition to deposition variables controlled by the experimenter, such as deposition rate and temperature, the step density of the substrate also has a significant effect on (4×4) growth. In areas of high local step density, growth on terraces is completely suppressed in favor of growth from steps. In contrast, in regions of medium and low step density, growth occurs both at steps and on terraces.

In summary, three key variables influence (4×4) growth as viewed in LEEM: temperature, deposition rate, and substrate step density. Under all growth parameters examined, the (4×4) phase was found to nucleate initially at step edges for coverages less than 0.1 ML. For sufficiently wide substrate terraces, subsequent (4×4) island growth on terraces was also observed. For island growth on terraces, the number density increases with deposition rate and decreases with deposition temperature. Too few data points were taken to compare the effects of substrate terrace width and deposition parameters on (4×4) island growth to theory, as was elegantly performed in a study of the growth on CaF$_2$ on Si(111) [35]. However, conclusions can be drawn as to the relative importance of terrace width, deposition rate, and temperature on the average domain size of (4×4) islands.

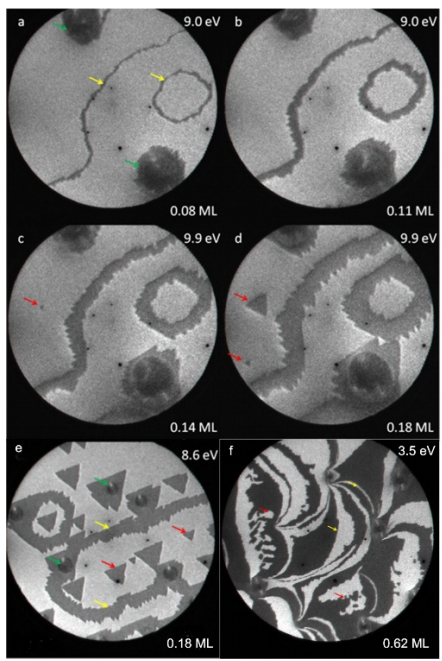

FIG. 4. LEEM images of the growth of (4×4) and $\sqrt{3}$ phases of Ag on Ge(111) at 410°C as a function of coverage. Colored arrows point at growth at three sites of the substrate: step edges (yellow arrows), terraces (red arrows), and at two large substrate defects (green arrows). Deposition rate is 0.017 ML/min. (a)-(d) Growth at low coverage shows (4×4) Ag (dark) on Ge(111) (bright). Field of view (FOV) = 5μm. (e) Another region with 0.18 ML coverage. Examples of substrate defects are marked with green arrows. FOV = 10μm. (f) Higher coverage of 0.62 ML, showing finger-like growth (red arrows) of bright √3 phase into darker (4×4) regions. FOV = 10 μm.

In terms of the controllable experimental parameters, temperature and deposition rate, temperature has the more dramatic effect on the number density of (4×4) islands, given the temperature range over which the (4×4) phase grows (~200-550°C) and given the range of deposition rates conveniently achievable with our Ag evaporator. Our experiments showed that a difference of 80-90°C in deposition temperature has a greater effect on island density than an order of magnitude difference in deposition rate.

Given a flat substrate, with wide terraces, the number density of Ag (4×4) islands is highly amenable to control. Large islands can be grown at high temperatures and low deposition rates. For example, Fig. 5 (e) depicts islands of ~1000 nm in width for deposition at T = 410°C, Θ/t = 0.017 ML/min, Θ = 0.18 ML. In contrast, small islands are grown by depositing at low temperature and high rate of deposition.

Approximate requirements for the step density necessary to grow islands of a desired size and number density can be estimated from our data. For temperature of 410°C and deposition rate of 0.017 ML/min, terraces greater than 2300 nm in width are necessary to nucleate (4×4) islands. We observed that a 70°C reduction in temperature results in approximately an order of magnitude reduction in the minimum terrace width required to grow Ag islands.

The effect of terrace width on growth was also observed for the higher density $\sqrt{3}$ phase, which nucleates after the completion of the (4×4) phase at 0.375 ML. While the nucleation of the $\sqrt{3}$ was observed to occur irrespective of the location of step edges, the steps were observed to form boundaries to $\sqrt{3}$ growth. FIG. 4 (f) shows $\sqrt{3}$ growth across a low step density sample at 410°C. Growth morphology across terraces is dendritic (red arrows) but bounded by step edges (yellow arrows).

The (4×4) growth begins at step edges for all experimental parameters examined. Whether any growth begins on terraces depends upon deposition temperature and deposition rate, as well as the local step density. Growth on terraces could be completely suppressed for samples of sufficiently high step density at a given deposition temperature. Alternatively, on samples of sufficiently low step density at a given temperature, (4×4) island growth was substantial. Growth at both steps and on terraces was characterized by faceted growth fronts.

While growth of the $\sqrt{3}$ phase was substantially different from that of the (4×4) phase, it was still strongly influenced by substrate step density. The $\sqrt{3}$ phase generally nucleated on terraces and grew dendritically away from the nucleation site until a step edge was encountered. At a given temperature, growth then proceeded in stripes down the lengths of terraces for samples of high step density, and islands had irregular shapes for samples of low step density.

Similar to previous STM studies [2,7,9], the (3×1) phase was observed in narrow spatial regions along (4×4) Ag domains in STM images. Although (3×1) spots are also apparent in LEED pictures, the domains are too small to resolve with LEEM. FIG. 5 shows LEED, LEEM, and STM measurements of the same sample, which had 0.59 ML of Ag deposited on Ge(111) at 330°C. At this coverage and deposition temperature, Ag on the surface has primarily $\sqrt{3}$ and (4×4) structures, as seen in LEEM (FIG. 5 (a)). However, LEED measurements show weak (3×1) spots, as well as strong (4×4) and $\sqrt{3}$ spots (FIG. 5 (b)). After depositing Ag while imaging with LEEM and LEED, the sample was cooled to room temperature without a significant change in either the LEEM image or the LEED pattern. The STM image shows (3×1) regions alongside edges of (4×4) domains, especially as small insets between (4×4) regions and small domains (<10 nm in size) of Ge(111)-c(2×8) that persist at this coverage (FIG. 5 (c)). The sizes of the (3×1) and (4×4) unit cells from our STM images agree with the dimensions reported in previous studies [2,7,9].

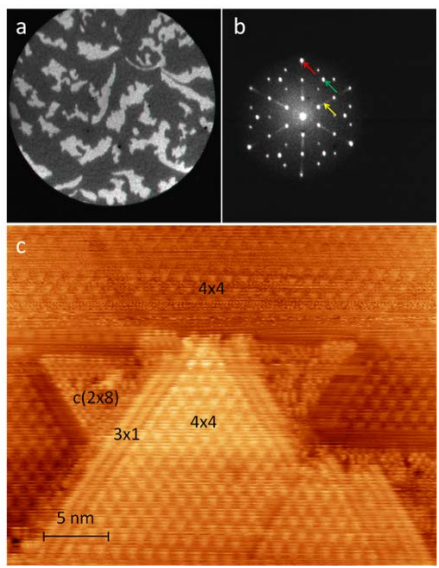

FIG. 5. Ag deposited on Ge(111) at 330°C while imaging with LEEM and LEED, followed by STM imaging at room temperature. Θ = 0.59 ML. (a) LEEM image showing regions with $\sqrt{3}$ (bright) and (4×4) (dark) structures. FOV = 5 µm, E = 4.0 eV. (b) LEED picture showing $\sqrt{3}$, (4×4), and faint (3×1) spots. For reference, several spots are indicated with colored arrows: $\sqrt{3}$ (green), (3×1) (yellow), and the [0, ¾] (4 x 4) spot (red). E = 8.1 eV. (c) Empty state STM image. Regions of (4×4), c(2×8), and (3×1) Ag are indicated. Sample bias = +1.0 V, tunneling current = 0.5 nA. 33.1 nm x 41.7 nm.

The (3×1) phase was observed to form at low coverage, before growth of the (4×4) or $\sqrt{3}$ phases. For deposition at 370°C, (3×1) was the only phase of Ag on the surface for coverage up to 0.1 ML. FIG. 6 shows LEED patterns for Ag deposited on Ge(111) up to ~1 ML at 370°C. The first Ag LEED spots observed are somewhat faint, elongated spots consistent with three rotational domains of Ag (3×1) (FIG. 6 (b)). As additional Ag is deposited, Ag (3×1) spots intensify as (4×4) (FIG. 6 (c)) and then $\sqrt{3}$ (FIG. 6 (d)) phases form. As the coverage approaches 1 ML, the (4×4) and (3×1) phases are replaced by the $\sqrt{3}$ phase on the surface (FIG. 6 (ef)). The (3×1) and (4×4) spots extinguish approximately simultaneously.

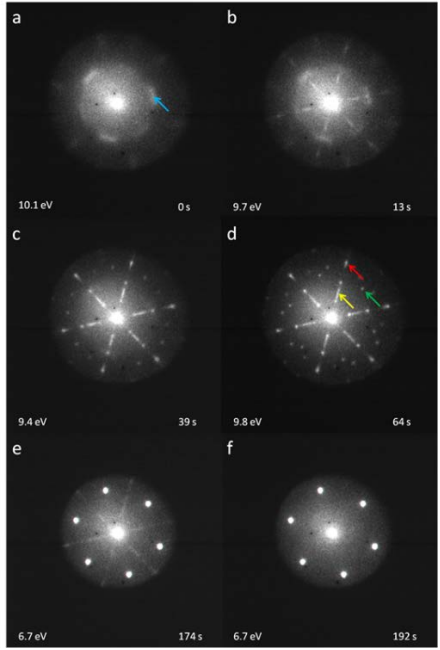

FIG. 6. LEED pictures of the deposition of Ag on Ge(111) at 370°C. Deposition rate = 0.3-0.4 ML/min, with seconds of Ag deposition indicated. Colored arrows in (d) identify LEED spots: Ag (3×1) (yellow), Ag (4×4) (red), and Ag $\sqrt{3}$ (green). The blue arrow in (a) points to one of the oblong diffraction features characteristic of Ge (2×1) (a) Ge (2×1). At this energy the primary Ge(111) LEED spots are not visible. (b) Ge (2×1) + Ag (3×1). (c) Ag (3 ×1) + Ag (4×4). (d) Ag (3×1) (yellow arrow) + Ag (4×4) (red arrow) + Ag $\sqrt{3}$ (green arrow). (e) $\sqrt{3}$ Ag + Ag (3×1) (faint) + Ag (4×4) (very faint). (f) $\sqrt{3}$ Ag.

We present an experimental phase diagram (FIG. 7), based on both the experimental data shown in this paper and additional data [36]. Note, however, that no attempt was made to include the effects of thermodynamics. In addition, the $\sqrt{3}$ phase which was observed in FIG. 1 is not shown in this phase diagram, because its appearance is hysteretic. Although for deposition below 200⁰C, it appears first, after heating to 370⁰C, it no longer recurs when the sample is cooled. Above 0.375 ML Ag coverage, LEEM and LEED images show that the $\sqrt{3}$ structure is very stable below 570⁰C. For this high coverage range, changing the temperature from 150⁰C to 575⁰C coverage resulted in an increase in the Ag $\sqrt{3}$ domain size with increasing temperature, as observed in LEEM.

For the region with T<300⁰C and coverage <0.375 ML, we have indicated the phases [c(2x8), 4x4, and 3x1] that appear to be stable when the sample is cooled to this temperature range.

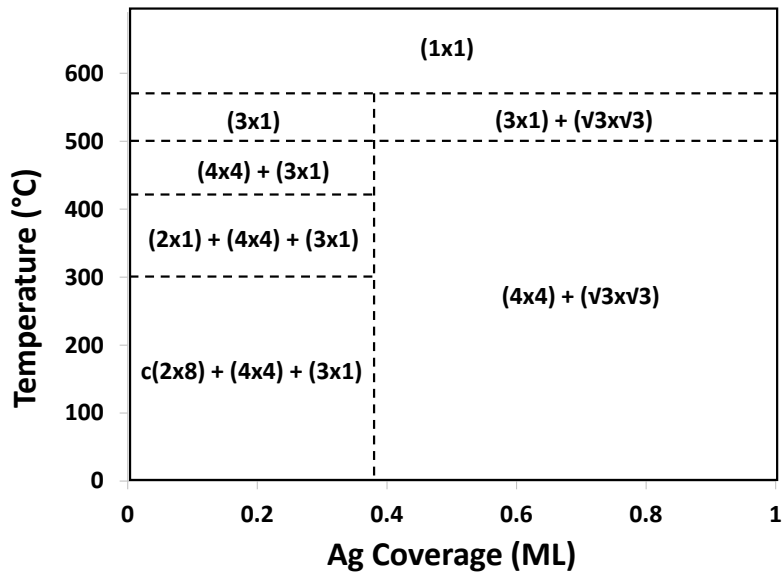

FIG. 7. Experimentally determined phase diagram for Ag/Ge(111).

The regions of the phase diagram showing the (3×1) phase differ somewhat from those in the phase diagram of Grozea [1], who observed this phase only for T>350⁰C. Instead, FIG. 7 shows this phase appears for a large range of lower temperatures. Grozea shows the (3×1) appearing in narrow temperature ranges above 350⁰C, but we see it occurring at somewhat higher temperatures. In other words, we observe the (3×1) over large regions of the phase diagram. While we observe (3×1) patterns easily in LEED pictures, the regions are not observed in LEEM real space images, either because they are too small or because they do not have sufficient contrast. The STM image in FIG. 5 (c) agrees in only showing (3×1) in a very small spatial region.

We also note that the (4×4) phase does not appear above 500$^0$C, in agreement with Grozea's phase diagram. At 575$^0$C, LEEM and LEED images showed that the (3×1) or the $\sqrt{3}$ structure disorder to a 1×1 or 3×1 phase. Ag then desorbs from the surface at 580$^0$C.

From the experimental work, we observed overlayer structures for a range of temperatures and coverages: Ge(111)-(2×1), Ge(111)-c(2× 8), Ag-($\sqrt{3} \times \sqrt{3}$)R30$^0$, Ag-(4×4), and Ag-(3×1). In the following sections, we provide details of our computational study to find atomic models of the observed phases, their electronic structure, their vibrational dynamics and their thermodynamic stability.

## IV. COMPUTATIONAL DETAILS

We perform DFT [21,22] based calculations using the Vienna Ab Initio Simulation Package (VASP) code [37]. The Perdew-Burke-Ernzerhof (PBE) [38] generalized gradient approximation (GGA) is used for the exchange-correlation (XC) functional. The pseudopotentials from the projected augmented wave (PAW) method [39,40] are used to treat the electron-ion interaction, and plane wave basis functions up to kinetic energy cutoff of 500 eV are used to expand the electron wave functions. A primitive unit cell of bulk Ge is constructed by taking two Ge atoms on a basis set located at (0, 0, 0) and (1/4, 1/4, 1/4) a, where 'a' is the lattice constant of face-centered cubic diamond structure. The 12×12×12 Monkhorst-Pack set of k-points is used for the Brillouin zone integration. All structure models are relaxed using the conjugate gradient algorithm with force and energy convergence criteria of 0.001 eV/Å and $10^{-5}$ eV, respectively.

The optimized lattice constant of bulk Ge after fitting the stabilized jellium equation of state [41] comes out to be 5.783 Å which is the overestimated value by 2.22 % from the experimental value of 5.657 Å [42]. Such an overestimation on using PBE is known to be due to the gradient correction on density that usually favors the more inhomogeneous system leading to elongation of bonds [43]. The value of bulk modulus is calculated to be 58.678 GPa which is the underestimates by 22.59 % from the experimental value of 75.8 GPa [44] The close prediction of bulk modulus using LDA XC in ref. [45] implies that the gradient correction on density can be attributed to the source of difference in value of bulk modulus. The electronic band structure obtained using PBE XC potential becomes metallic with no band gap as shown in the FIG. 8 (a) with green curves which is in disagreement with the experimental band gap of 0.74 eV at low tempearature and 0.66 eV at 300 K [46]. Such an underestimation of band gaps of semiconductors when using PBE and LDA XC in the DFT approach is well known and is attributed to simplying assumptions made about the discontinuity of the functional derivative of the exchange-correlation [47,48]. Over the years several efforts have been made to overcome this deficiency of DFT and methods such as GW,DFT with hybrid functionals, linear combination of atomic orbitals (LCAO) formalism with GGA potential and **k.p** method have successfully reproduced the band gap of bulk

Ge [49-52]. These methods are, however, computationally demanding for multiple calculations each requiring super cells consisting of few hundred atoms, as is the case here. Note that our interest is in calculations of the electronic structure and vibrational dynamics of surface systems which need upto 178 atoms in the super cell.

On the other hand, it has been shown that the use of DFT+U method [53] with isotropic on-site effective interaction $U_{effective}$ = -J can reproduce the experimental band gap and elastic constants of bulk Ge [54,55]. We use the same approach of tuning the value of $U_{effective}$ to match the experimental band gap in this study and find it to be computationally feasible for our purposes. For $U_{effective}$ = -2.94 eV that enhances the strength of the effective on-site interaction between p electrons, close to its reported values [54-56], we get an indirect band gap ($E_g^{\Gamma-L}$) of 0.66 eV (FIG. 8). In FIG. 8 (a), one can see that the correction to the energy of p electrons leading to their localization ($U_{effective}$ < 0) results in increase in eigen-energy of the sp hybridized bands at non-zero value of k (most significantly on the lowest conduction band) and adds contribution of p states near the top of the valence band near the Γ point. Since p-states have zero probability at the atomic sites, pure p-states are not affected by $U_{effective}$ as shown by overlapping of green and red curves at the top of valence band. In accordance with group theory, there are pure states at Γ point: triply degenerate p-states at top of valence band and the single s-state above it. At this set up, the lattice constant and bulk modulus of Ge obtained using the jellium equation of state are 5.616 Å and 80.577 GPa, respectively which are 0.72 % and 7.3 % off from experimental value of 5.657 Å and 75.8 GPa reported in refs. [42] and [44], respectively. Since p orbitals of higher energy are available continuousy at and near the top valence band in the band structure from DFT+U approach (which are absent in using DFT with PBE XC) (see FIG. 8 ), that form strong bond resulting into relatively small lattice constant in DFT+U approach. The distance to first four neighbors is 2.432 Å and angles in the tetrahedral coordination are $109.5^0$ implying $sp^3$ hybridized orbitals of bulk Ge atom.

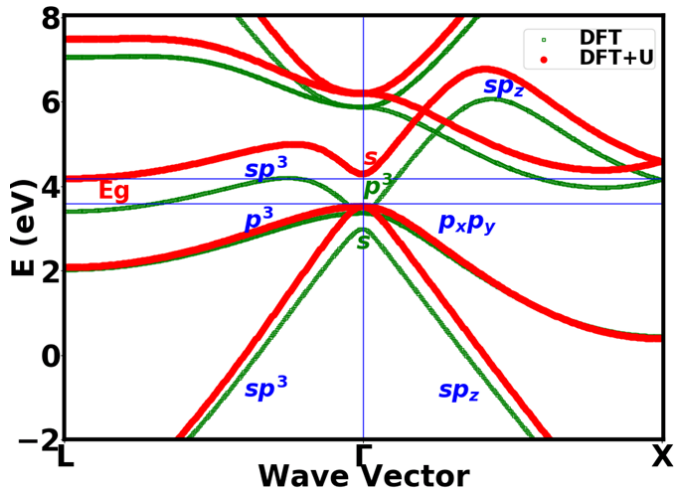

FIG. 8. Electronic band structure of bulk Ge calculated using the DFT (thin green curve) and the DFT + U ($U_{effective}$ = -2.94 eV) approach (thick red curve) alongwith the labeling of contributing orbitals on each band where for e.g., $p^3$ refers the contribution of all three p orbitals.

For surface construction using the slab models, we use a supercell having 15 Å vacuum in the normal direction to the slab in order to minimize the interactions between periodic images. The 12×12×1 Monkhorst-Pack k-point mesh for the (1×1) surface unit cell is scaled in calculation of surface phases with different periodicity to have the same numerical accuracy. The surface free energy at temperature T, $\gamma(T)$, is used to compare energetics of surfaces with different numbers of atoms and is calculated as

$$\gamma(T) = \frac{1}{2A} [\, G_{slab}^{Total}(T) - \sum_i \mu_i(T) N_i \,], \qquad (1)$$

where $G_{slab}^{Total}(T)$ is the Gibbs free energy of the slab, $\mu_i(T)$ and $N_i$ are the chemical potential and the number of atoms of type 'i' in the system, respectively, 'A' is the area of surface unit cell and the factor (½) is due to 2 identical surfaces on slab.

In equation (1), $G_{slab}^{Total}(T)$ is calculated as the sum of the total electronic internal energy of slab supercell obtained from DFT calculation ($E_{slab}^{Total}$) and the temperature dependent phonon Helmholtz free energy ($F_{slab}^{vib}(T)$) calculated using

$$F_{slab}^{vib}(T) = \int_0^\infty [\frac{\hbar\omega}{2} + k_B T \ln(1 - e^{-\frac{\hbar\omega}{k_B T}})] \, \rho(\omega) \, d\omega, \qquad (2)$$

in which the phonon frequencies $'\omega'$ are calculated by using the finite difference method and supercell approach as implemented in phonopy package [57]. To calculate force constant, atoms in $4 \times 4 \times 4$ supercell of bulk Ge (for which phonon density of states converge) are displaced from relaxed position with amplitude of 0.01 Å. Phonon density of states $\rho(\omega)$ is calculated by sampling the Brillouin zone with $90 \times 90 \times 90$ q-point mesh for bulk Ge with a smearing width of 0.05 THz and the points are scaled for different surface systems to have the same numerical accuracy. The plot of phonon dispersion and the corresponding density of states of bulk Ge are shown in FIG. 9 (a) and (b), respectively. The plots are in excellent agreement with the reported theoretical calculation [58] using density functional perturbation theory (DFPT) approach and experimental data [59,60].

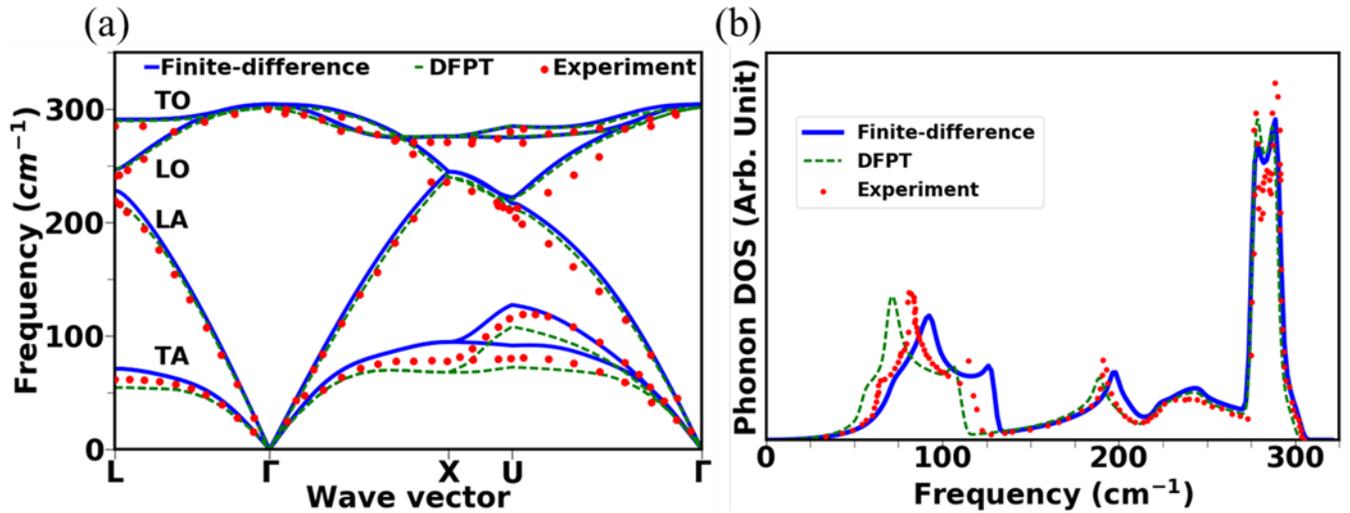

FIG. 9. (a) The phonon dispersion and (b) the phonon density of states, of bulk Ge obtained using finite difference method used in this study (blue curve) and comparison to that obtained from density functional perturbation theory based calculation [58] (green dotted curve) and experimental result: phonon dispersion from ref. [59] and phonon DOS from ref. [60] (filled red dots).

The chemical potential of Ge ($\mu_{Ge}$ (T)) in equation (1) is taken as the energy of a bulk Ge atom ($E_{Ge}^{bulk}$) which corresponds to the thermodynamic equilibrium between surface and bulk Ge atoms. To calculate $\mu_{Ag}$ (T) of Ag adatom, we have to approximate a reference from where the adatoms come to make the ordered structure. Since the Ag structures considered in this study are planar, their reservoir is approximated as the system with far apart separated Ag adatoms on the Ge(111) surface at which they behave like 2D-non-interacting gas that interact with substrate. Under the approximation, which we mimic by taking one Ag adatom on 6 × 6 surface unit of Ge(111), $\mu_{Ag}$ (T) is given by (see also ref. [61]):

$$\mu_{Ag}(T) = \varepsilon_{Ag} - \ln(\frac{2\pi mkT}{h^2} \frac{A}{N}), \qquad (3)$$

where $\varepsilon_{Ag}$ is the binding energy of an isolated Ag adatom on the relaxed Ag/Ge(111) system, 'A' is the surface area on which N = 1 Ag adatom is adsorbed. Once $\mu_{Ag}(T)$ of Ag is substituted in equation (1) for an overlayer, one can get Ag coverage (u) and temperature (T) dependent surface free energy $\gamma$ (T, u) of a phase.

For a given surface structure under consideration, its atomic geometry is taken to be the one that has the lowest $\gamma(0)$ among the various configurations considered with adatoms on different probable adsorption sites either on the intact double layer, or on the missing top layer Ge(111), since reported models of ordered structures of the Ge(111) surface or that with Ag overlayer display one of these terminations. The temperature (T) and coverage (u) dependent surface free energy $\gamma(T, u)$ of the energetically favored configuration is then calculated. If the value of $\gamma(T, u)$ of a thermodynamically stable phase is lower than that of other phases with the same number of atoms for the entire temperature range, the former is favored and so $\gamma(T, u)$ of that phase is taken for phase-diagram calculations. The $\gamma(T, u)$ of individual phases are combined to find $\gamma(T, u)$ of an arbitrary coverage and temperature numerically (section 5.8 for details). An isolated phase or coexistence of considered phases that has the minimum $\gamma(T, u)$ is then decided to be the phase at the given temperature and coverage condition. Such a numerical approach overcomes the computationally very intensive first-principles calculation of a single model of coexisting overlayer structure.

## V. RESULTS

We discuss the geometrical features of the energetically favored atomic models of unreconstructed Ge(111), reconstructed Ge(111) and of systems having Ag adatoms on the Ge(111) surface. The vibrational structure of model of each phase is presented as a measure of thermodynamic stability and its electronic structure is discussed to understand the formation of the phase. A surface phase diagram is constructed by taking the surface free energies of the dynamically stable atomic model of phases.

### A. Unreconstructed Ge(111)–1×1

The atomic model of the unreconstructed Ge(111)-1×1 surface formed by stacking bulk-like structure along [111] direction is shown in FIG. 10 (a) (side view) and (b) (top view). This is an intact double layer (IDL) surface structure in which the surface Ge atoms are coordinated with three Ge atoms on layer below at the nearest neighbor separation, different from another possible missing top layer (MTL) surface in which such coordination is one with atom directly below it. An ABC stacking of double layers is formed on both type of surface termination. The surface energy of the intact double layer (IDL) structure is 85.88 meV/Å$^2$, smaller than 150.07 meV/Å$^2$ (energy higher by 5.260 eV) of missing top layer (MTL) surface termination and 108.24 meV/Å$^2$ (energy higher by 1.832 eV) of reconstructed MTL (MLR) surface in which the surface Ge atoms of MTL termination form equilateral triangle.

In the side view of the IDL model of Ge(111) shown in FIG. 10 (a), the bond length between atoms on top surface layer (1$^{st}$) and a layer below (2$^{nd}$ layer) ($d_{12}$) is 2.402 Å and other similarly defined interlayers bond lengths are $d_{23}$=2.468 Å, $d_{34}$=2.428 Å, $d_{45}$=2.442 Å and $d_{56}$=2.432 Å (bulk value). In comparison to the bulk position of the atoms, the height of an atom in layer 1 is relaxed inward by 0.065 Å, in layer 2 by 0.034 Å outward, in layer 3 by 0.003 Å outward, and in layer 4 by 0.013 Å outward. So, the 1$^{st}$ and 2$^{nd}$ layer atoms whose s and pz orbitals contribute on the dangling bond (significantly more of 1$^{st}$ layer atom, see FIG. 11 (b)) relaxes more than inner layer atoms. Upon relaxation of the Ge(111) surface, the contribution of s and pz orbitals on dangling bond decreases slightly and the small contribution of px and py orbitals on the bond that appeared on unrelaxd surface vanishes. On the top view of the surface shown in FIG. 10 (b), atoms from three different layers are exposed. The distance between Ge atoms on the surface layer is 3.971 Å which is the second nearest neighbor distance in bulk.

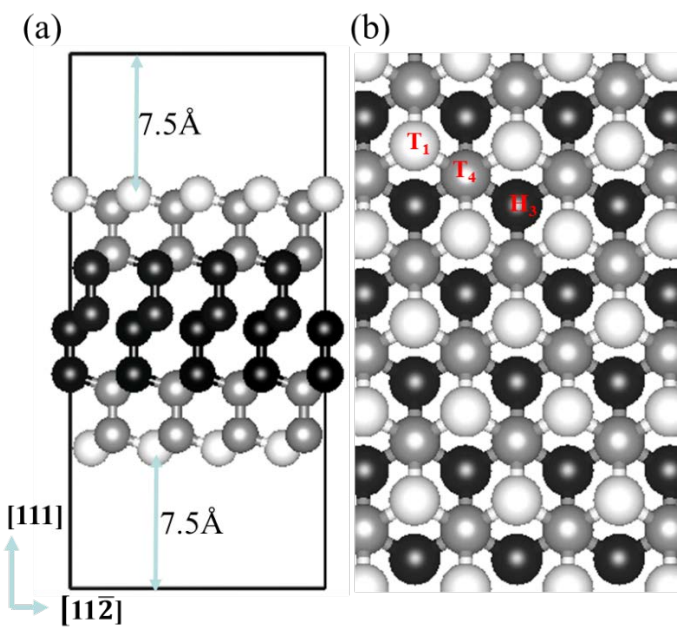

FIG. 10. (a) Side view of the repeated simulation supercell of the atomic model of the Ge(111)-1×1 phase. (b) Top view of the phase that exposes 3 different on-top adsorption sites: above first layer atom ($T_1$) (above white sphere), above four-fold coordinated second layer atom on three-fold hollow site ($T_4$) (above graysphere) and above fourth layer atom on three-fold hollow ($H_3$) site (above black sphere).

The electronic charge density distribution and the orbital projected density of states of atoms on the top surface layer and on the second layer of the Ge(111) surface are shown in FIG. 11 (a) and (b), respectively. In FIG. 11 (a), there are three red colored dumb bell shape high electron density regions between each top layer Ge atom (position corresponds to center of small blue circles) and three Ge atoms on layer below (position corresponds to center of green contour) that represent their covalent bonding and three-fold symmetric coordination of surface atoms.

Geometrically, the distance from top layer Ge atom to Ge atoms on $2^{nd}$ layer is 2.402 Å, less than the bulk nearest neighbor distance of 2.432 Å. The bonded atoms subtend an angle of $111.5^0$ ( angle by the centers of the green contours at the center of blue sphere), slightly different from that of $109.5^0$ at bulk structure, at the top layer Ge atom due to change in interlayer distance discussed previously. In FIG. 11 (b), the dangling bond is shown to be localized only on the top surface layer atom (atom-1) and it is of hybridized s and $p_z$ orbitals . Since the atom directly below the top layer surface atom is after two double layers, the pz bond is not saturated. Note that z direction is taken to be perpendicular to the surface i.e., the direction perpendicular to the plane containing white spheres in FIG. 10 (a) is z-direction. The second neighbor distance separation of surface atoms imply that the surface dangling bonds interact through second-neighbor.

The phonon dispersion of Ge(111)-(1x1) shown in FIG. 11 (c) points to its propensity for reconstruction. The two acoustic modes, one longitudinal and another transverse, have imaginary frequency along Γ-M or Γ-K directions, implying that the surface undergoes reconstruction. In fact there are two different observed phases of the reconstructed Ge(111) surface, as discussed in section 3: Ge(111)-2×1 formed on cleavage [62] [63-67] and Ge adatom decorated Ge(111)–c(2×8) phase [68,69] .

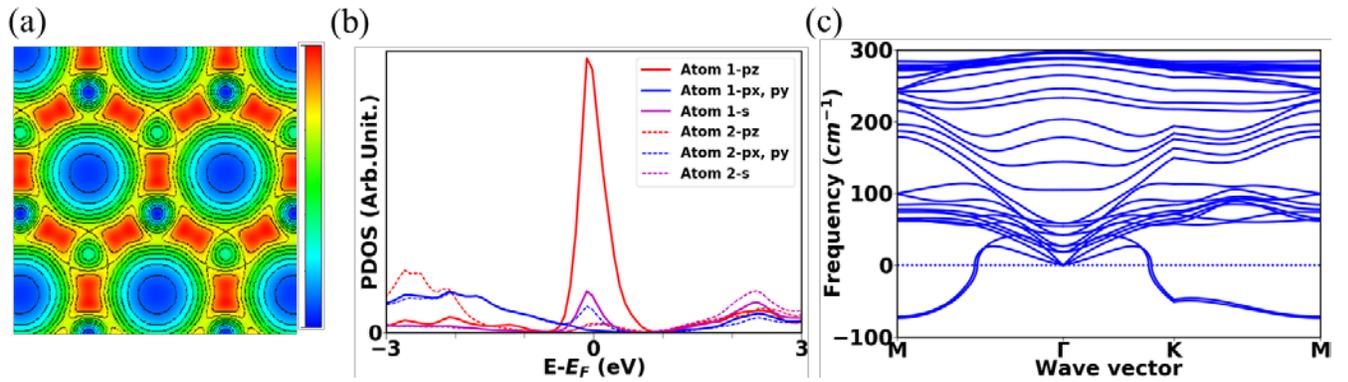

FIG. 11. (a) Electron-density distribution on a plane between the top Ge layer and 2$^{nd}$ Ge layer from top. Contours are plotted up to 0.08 $A_0^{-3}$ with the interval of 0.008 $A_0^{-3}$. The center of small blue, green, and large blue contours correspond to the atoms such that sites below them are assigned as $T_1$, $T_4$ and $H_3$, respectively. The red colored dumb-bell contours corresponding to high density represent sp$^3$ type covalent bond between Ge atoms. (b) The projected density of states (PDOS) along ps and s orbitals of Ge atom on top surface layer (atom-1) and 2$^{nd}$ layer (atom-2). Note that the z-direction in the PDOS plot in all figures is taken perpendicular to the surface. (c) Phonon dispersion, of Ge(111)-1× 1 system.

## B. Reconstructed Ge(111)-2×1

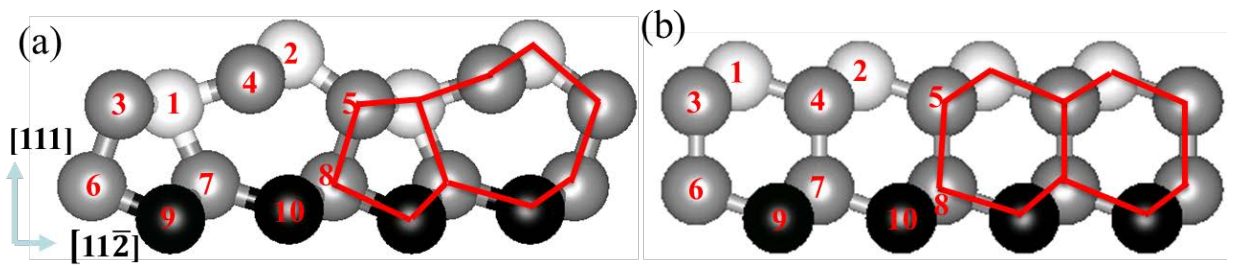

FIG. 12. Side view of the repeated unit cell (labeled atoms, except 5 and 8, belong to a unit cell) of (a) the reconstructed Ge(111)-2×1 phase, and (b) the unreconstructed Ge(111)-1×1 system for reference. The white, gray, and black spheres represent $T_1$, $T_4$ and $H_3$ adsorption sites, respectively. The numbering on atoms is to facilitate a comparison of atomic positions.

In 1961, Haneman [62] observed diffraction pattern with half-order beams on the (111) surfaces of diamond structure crystals including Ge. Since such a pattern can be obtained from surface with atomic periodicity double of the spacing of atoms in bulk-like terminated (111) planes, a buckled model was purposed with alternating outer atoms raised and lowered relative to an ideal bulk termination[62].This model is different from the Pandey Π-bonded chain model for Si(111)-2×1 [63,70,71] and C (111)-2×1 [72] in which atoms in the Π-chain are bonded with two other surface atoms and lie in the same horizontal plane without buckling or tilting of the chain atoms. Various theoretical [64-66] and experimental [67] studies confirm that the Ge(111) surface has modified Pandey Π-bonded chain structure in which the atoms in Π chain are tilted so that one of the atoms in the tilted structure becomes outer and the other inner. Similar structures are reported for Si(111)-2×1 too [73-76]. There are two tilted buckling geometry depending on whether one or the other of the Π-chain atoms is outermost: one in which the tilting is reverse to the atomic orientation in the 3rd layer from surface and is called negative buckling (of atoms 4 and 2 with that of 7 and 10 in FIG. 12 (a)) and the other isomer in which tilting is parallel to orientation of the 3rd layer atoms and is called positive buckling (not shown).

In our calculation, the value of $\gamma$ (0 K) of negative buckled structure (side view in FIG. 12 (a)) is 72.32 meV/Å$^2$ which is lower than that of positive buckled structure by 0.866 meV/Å$^2$ (energy difference by 0.047 eV per unit cell or 0.0023 eV per atom). The untitled Pandey model transforms into the negatively tilted model upon relaxation. Although the energy difference between two types of tilted structure is small, the favored tilting in our calculation is different from positive buckled structure in other studies [64,77]. In comparison to unreconstructed Ge(111)-1×1 phase, the value of $\gamma$ (0 K) of the negative tilted reconstructed phase is lower by 13.37 meV/Å$^2$ which implies that the reconstruction is energetically favored. The relative energy gain with respect to the unreconstructed surface is 0.36 eV/(1×1) surface cell, in close agreement with reported value of 0.334 eV [64]. The side view of the unreconstructed Ge(111)-1×1 phase is shown in FIG. 12 (b) for reference. Although all of our unit cell parameters are scaled with the bulk lattice constant, a test calculation shows increase of buckling when the cell volume is not contrained as indicated by 6.8%, 1.4%, 4% and 0% change in the height between atoms labelled 1 and 4, and 2, 2 and 5, and 3 and 1, respectively.

Geometrically, one of the surface layer atom of Ge(111)-1×1 (atom 1 in FIG. 12 (b)) shifts in the [$\bar{1}\bar{1}\bar{1}$] direction (Fig. 12 (a)) and stabilizes at 0.1 Å below atom 3 of second layer. The atom labeled 1 bonds with four atoms labeled 3 , 3' (next to 3 on surface), 4 and 7 at distance 2.452 Å, 2.452 Å, 2.411 Å and 2.524 Å, respectively subtending angles 3-1-4 of 124$^0$ and 3-1-3' of 108.2$^0$. The decrease in electronic energy by the formation of bonding requires significant bond distortion. To minimize the energy corresponding to bond distortion, atom 1 aligns at the same height with atom 3 of second layer and atom 3 undergoes lateral displacement towards atom 1 by 0.824 Å. Both atoms 3 and 1 are bonded with four atoms at distance close to the bulk bond lengths and so have bulk-like coordination on the surface. Atom 7 whose bond with the atom on top of it is broken is also bonded with four atoms, three on the same $2^{nd}$ layer and one on the $1^{st}$ layer. Atom 4 in FIG. 12 (a) bonds with three atoms: with tilted atom 2 and 2' (behind 2 on surface) at distance 2.387 Å and with atom 1 at 2.411 Å, all being less than the bulk bond length of 2.432 Å and subtending angles 2-4-2' of $112.6^0$, 1-4-2 and 1-4-2' of $122.5^0$. The atom 2 bonds with three atoms: atom 4 and 4' at 2.387 Å subtending angle 4-2-4' of 112.6 and with atom 5 at distance 2.470 Å subtending angle 5-2-4 of $102.4^0$. As a result of buckling or tilting, rather than hexamer structure, the system forms five-atom and seven-atom rings. and the dangling bonds disappear.

Since both the dimerization and buckling decrease the electronic energy, the reconstructed surface is energetically favored over the clean unreconstructed surface, thanks to rearrangements of bonds in the topmost two atomic layers. Its model has five-fold and seven-fold rings shown in the side view in FIG. 12 (a), in agreement with the buckled Pandey chain model [77]. The height difference (buckling) between atoms 1 and 4 in the reconstructed structure is 0.712 Å, and that between atoms 4 and 2 is 0.707 Å which is in close agreement with the reported value of 0.78 Å [64]. As a result of reconstruction, there is a reduction of the states near the Fermi level which form dangling bond in the unreconstructed Ge(111) surface, as depicted in the PDOS plot of the phase in FIG. 13 (a). There is a small non-zero value of the projection of density of states on $p_z$ orbital of atoms labeled 1 and 4 in FIG. 12 (a) at Fermi level, as seen in Fig. 13 (a). The phonon density of states of the phase shown in FIG. 13 (b) displays no soft modes, implying its thermodynamic stability. The peak about wavenumber 50 cm$^{-1}$ in FIG. 13 (b) corresponds to the vertical motion of the atom labeled 2 in FIG. 12 (a) in [$\bar{1}\,\bar{1}\,\bar{1}$] direction.

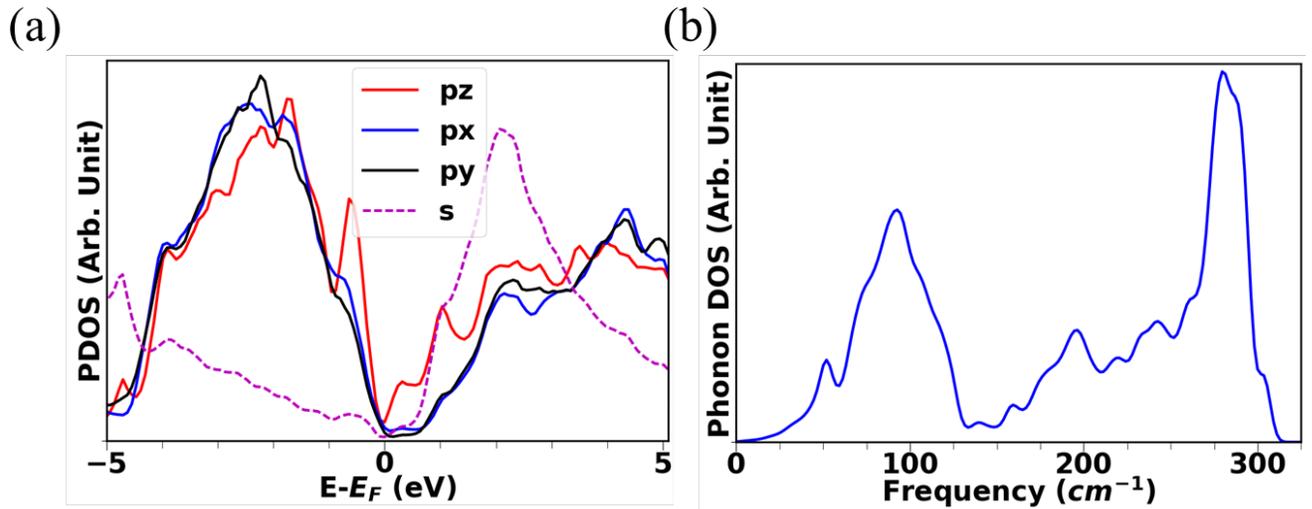

FIG. 13. (a) Projected density of states (PDOS) on s- and p- orbitals, (b) phonon density of states, of the Ge(111)-2×1 phase.

C. Reconstructed Ge(111)-c(2×8)

For an atomic model of the Ge(111)-c(2×8) phase, we turn to those previously proposed [69,78-81] in which there is one Ge adatom for every four surface atoms in a way such that ¾ of top layer dangling bonds of the Ge(111)-1×1 surface are saturated. After considering a number of scenarios for the position of the adatom, the model shown in Fig. 14 (a) with adatoms adsorbed on T$_4$ sites is found to be the energetically most favored configuration with γ (0) = 71.24 meV/Å$^2$, close to reported value of 68.62 meV/Å$^2$ [82]. The surface energy of the configuration is 8.098

meV/Å$^2$ lower than that with adatoms on the H$_3$ sites (21 meV/atom in terms of total energy), while the T$_1$ sites is not at all favored.

In the model, the distances between successive Ge adatoms in the same row along the longer side of the surface unit cell, shown in FIG. 14 (a), is 7.915 Å and 6.91 Å along the shorted side. The surface Ge atoms around an adatom show displacement different from 3.971 Å found on Ge(111): three of them from top Ge(111) surface layer are at 3.771 Å (pulled towards adatom) and other three atoms from two layer below are at 4.126 Å (pushed away from adatom). A Ge adatom pushes down a Ge atom below it by 0.64 Å with respect to other atoms on the same layer and the pushed atom similarly does for 3$^{rd}$ layer atom below it by 0.52 Å. Such an observation of pulling up of adatom at the T$_4$ site and pushing down the atom below adatom is also observed on Si(111) –(7×7) surface and is attributed to the rehybridization of the sp$^3$ back bonds [83]. An additional surface feature is the pop up of "rest" atoms, the three-fold coordinated Ge surface atoms that are not bonded with adatoms, by 0.33 Å in our study. Such popping up of rest atoms is also observed in STM [79] and helium atom scattering [84]) experiments. The buckling of 0.025 Å between rest atoms in the (2×2) and c (2×4) sub-units of the c(2×8) surface unit is also in agreement with STM observations [79]. Our calculations do not show buckling between adatoms, in agreement with previous work [69]. The structural asymmetries in rest atoms but not on adatoms has been explained [69] in terms of the presence of small excess electron charge in rest atoms but not on adatoms.

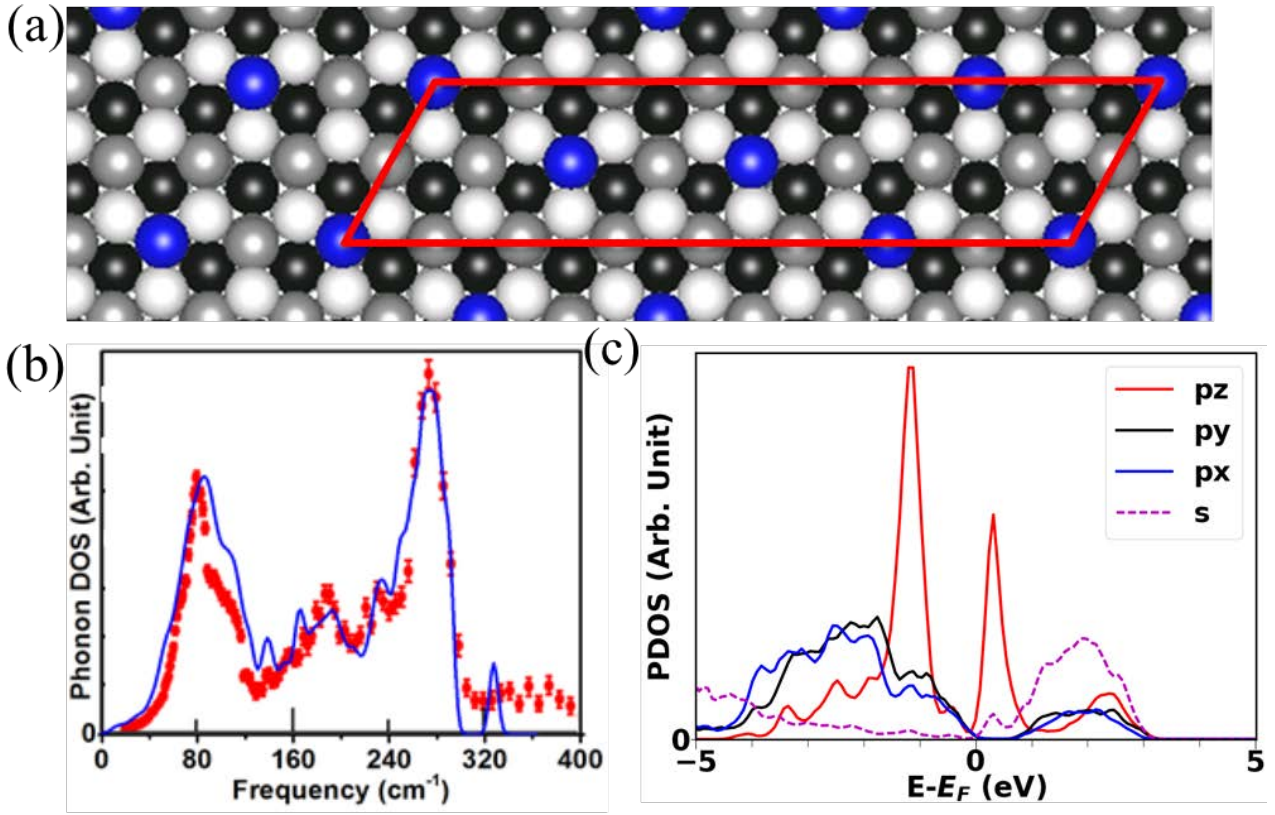

FIG. 14. (a) Top view of the Ge(111)–c(2×8) surface with Ge adatoms (blue spheres) adsorbed above gray spheres ($T_4$ site). (b) Comparison of phonon density of states (continuous blue curve) with experimental measurement [85] (discrete red points). (c) Projected density of states (PDOS) on s- and p- orbitals.

In FIG. 14 (b), comparison of the calculated phonon density of states with available experimental data [85] shows good agreement. The high frequency mode around 320 cm$^{-1}$ results from the short bond of length (2.364 Å) between the atom directly below the adatom and the 3$^{rd}$ atom just below it. This feature is a signature of adatom adsorption on the $T_4$ site and does not exist when adatom adsorbs on the hollow $H_3$ site since there are no such stretched bonds. The projected density states of the phase shown in FIG. 14 (c) is metallic due to the presence of dangling bond on adatoms and rest atoms.

An indication of the relative stability of the atomic models of unreconstructed and reconstructed Ge(111) surfaces is obtained through comparison of their surface free energy, $\gamma(T)$, in FIG. 15. For the temperature range of interest it shows that $\gamma(T)$ of the unreconstructed Ge(111)-(1×1) phase is higher than that of both reconstructed surfaces, and that the Ge(111)-c(2×8) phase is slightly favored over the (2×1). This relative stability of the of c(2×8) phase over the (2×1) can be traced to the energy gain due to shift of occupied bands of the rest atom into the projected bulk valence bands [86]. In short, without metal adsorption, the Ge adatom decorated Ge(111)-c(2×8) structure is the energetically favored and thermodynamically stable phase, in agreement with experimental observations of this phase [68,79,80,84,87-89].

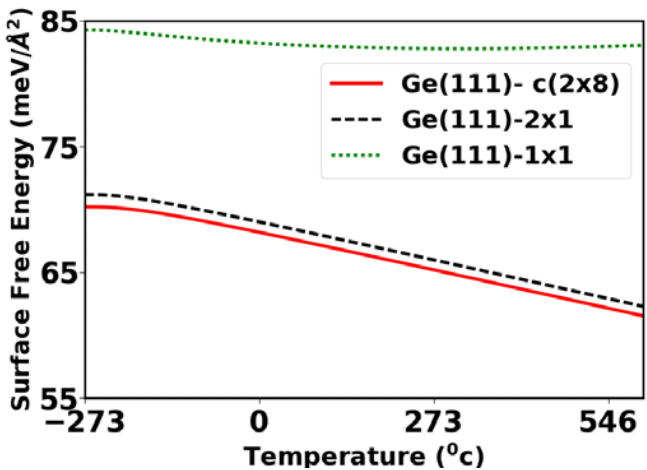

FIG. 15. Variation of surface free energy with temperature of the unreconstructed and reconstructed phases of Ge(111).

## D. Ge(111)-Ag (1×1)

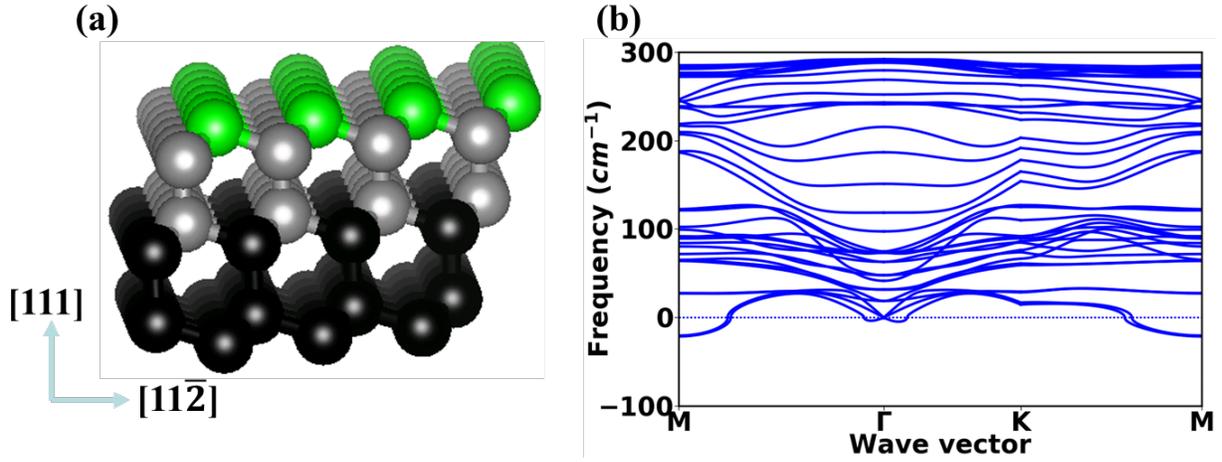

FIG. 16. (a) Side view of the repeated atomic model of the Ge(111)-Ag (1x1) phase in which the green spheres represent Ag adatoms, and gray and black spheres represent the $T_4$ and $H_3$ adsorption sites, respectively. (b) Phonon dispersion of the Ge(111)–Ag (1×1) phase.

Our calculations find the atomic model in which Ag adatoms adsorb on the missing top layer (MTL) of Ge(111) at 1 ML coverage (see FIG. 16 (a)) to be energetically the most favored configuration over those in which the Ag adatom adsorbs on the $T_4$, $T_1$, and $H_3$ sites of the intact double layer surface by 16, 6.57 and 5.34 meV/Å$^2$, respectively. The height of the Ag atom over the nearby Ge atom are 0.924 Å, 1.529 Å, 1.732 Å and 2.466 Å on MTL, $H_3$, $T_4$ and $T_1$ adsorption configurations, respectively, showing that Ag is closest to surface in the MLR configuration. Additionally, in the MLR structure, the Ag adatom adsorbs at distance 2.472 Å from three surface Ge atoms, a value exactly equal to the distance in $\sqrt{3}$ structure discussed below. The interlayer Ge distances are $d_{23}$ (between 2$^{nd}$ and 3$^{rd}$ layer from top) = 2.417 Å, $d_{34}$ = 2.435 Å, $d_{45}$ = 2.438 Å and $d_{56}$ = 2.432 Å (bulk value). The comparison of the bond lengths with unreconstructed Ge(111)-1× 1 system shows that on Ag adsorption, the length of bonds between an atom and another atom on top of it are decreased while that of others increase.

The phonon dispersion of the MLT phase shown in FIG. 16 (b) displays modes with negative frequencies implying dynamic instability of the structure. As in the case of clean Ge(111) surface, these are the two acoustic, one longitudinal and another transverse modes. Following the negative peak at M (1/2, 0, 0) point of Brillouin zone, the structure is deformed along the Γ – M direction to break symmetry. To accommodate the deformation in calculations with periodic boundary condition, the unit cell is repeated 2× 2 × 1 times. The symmetry broken structure is then found to have lower surface energy than the initial 1× 1 structure. In the deformed structure, some Ag adatoms and some Ge atoms are closer in comparison to unreconstructed structure. Three of the Ag atoms form planar equilateral triangle centered about the $T_4$ site so that the Ag-Ag distance is 4.23 Å (which is 3.971 Å in the undeformed structure) and the height of the Ag layer from the Ge layer is 0.7 Å (which is 0.924 Å in the non-deformed structure), close to the value 4.56 Å and 0.727 Å of the same quantity in the $\sqrt{3}$ structure discussed below. The fourth Ag atom pops up by 0.2 Å (which is not case in undeformed structure) as compared to the other Ag atoms and is at the center of the three surface Ge atoms that form an equilateral triangle with side length of 4.42 Å (which is 3.971 Å in the undeformed structure). The distance of the fourth Ag atom from the closest lying Ag adatom is 2.855 Å, closer than the Ag bulk bond length of 2.932 Å. As a result of the above transformation the Ag adtoms centered about the $T_4$ site get further spaced from each other and end up closeer to the Ag atom which is at center of Ge atoms. This instability in the Ag (1 × 1) ordered structure forbids isolated existence of the ordered phase: a result consistent with experimental observations of the phase at temperatures beyond $400^0$ C or in the disordered phase without definite coverage [1,90].

### E. Ge(111)-Ag (3×1)

Hammar et al. [2] identified the existence of the (3×1) phase of Ag at the boundary of the (4×4) phase grown on the Ge(111)–c(2×8) surface with atomic geometry similar to that of this very structure on the Si(111) surface. Previous to the observation, the (3×1) phase was shown to form when alkali metals adsorb on the Ge(111) surface[91] and Ag on the Si(111) surface [18]. With the establishment of $\frac{1}{3}$ ML of the metal coverage [10], missing top layer or Seiwartz model [10-12], extended Pandey model [12,13] and honeycomb chain-channel (HCC) model [14-17,92] have been proposed on the Si(111) surface. For Ag/Ge(111) system, Grozea et al. [93] reported the existence of the (3×1) phase at large region at $\frac{1}{3}$ ML and Mullet et al. [94] observed the (3×1) phase on substantial region along with $\sqrt{3}$ and (4×4) phases especially at high temperature. The coverage and temperature regime in which the phase is observed as an isolated or in coexisting form has already discussed in sec. 3. However, despite these and related studies and debate, a detailed, microscopic understanding of the Ge(111)-Ag (3x1) phase from first principles is lacking. We present here results of such an investigation which points to the honeycomb chain-channel (HCC) model whose side and top view are shown in FIG. 17 (a) and (b), respectively, as the energetically favored structure with surface energy lower by 11.49 meV/Å$^2$ than that of the Seiwartz-chain model and by 25.26 meV/Å$^2$ over the extended-Pandey-chain model. The lack of soft modes in the phonon density of states presented in FIG. 17 (c) lead us to conclude that the HCC structure is thermodynamically stable.

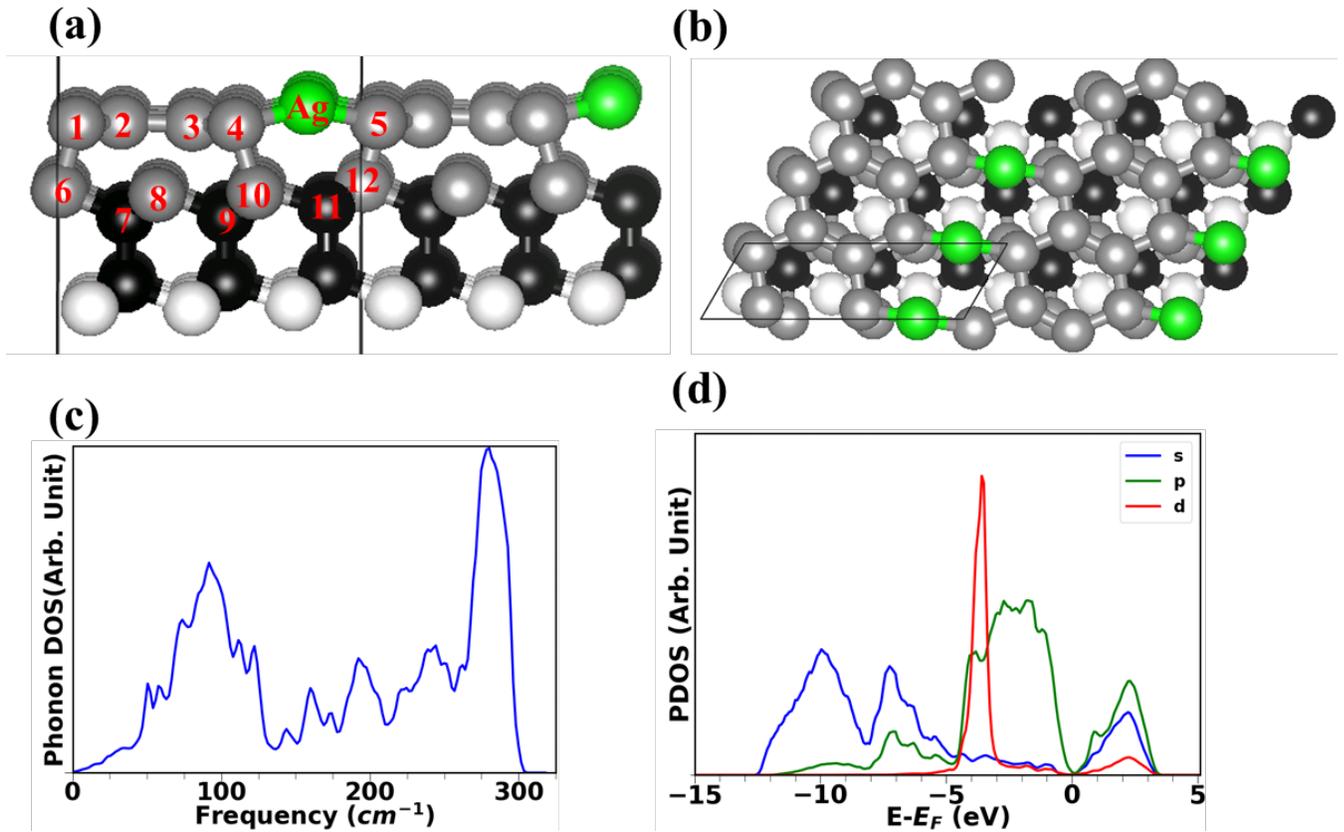

FIG. 17. (a) Side view of the equilibrium geometry of the honeycomb chain-channel (HCC) model of the Ge(111)-Ag (3×1) phase in which the green spheres are the Ag adatoms and the gray, black and white spheres represent Ge atoms in different double layers. The vertical lines enclose the supercell laterally. (b) Top view. (c) Phonon density of states. (d) Projected density of electronic states.

The surface unit of the HCC model has four inequivalent Ge atoms (labeled 1, 2, 3 and 4 in FIG. 17 (a)) forming a honeycomb chain that runs along [1 $\bar{1}$ 0] direction in a plane parallel to the bulk terminated (111) substrate and metal atom in channel. The maximum height variation among the Ge atoms in the honeycomb-chain is 0.096 Å. The Ag atom in the channel is at nearly equal distance from the closeby Ge atoms: 2.646 Å from Ge (4) and 2.659 Å from Ge (5) in FIG. 17 (a). This distance is slightly more than the Ag-Ge distance of 2.472 Å in (1×1) and $\sqrt{3}$ structures. The Ag-Ag distance in channel is 3.971 Å. The outer atoms 1 and 4 are coordinated with four atoms, three Ge atoms and one Ag atom, while the inner atoms 2 and 3 are coordinated with three Ge atoms since d (Ge(2)-Ge(8)) =2.766 Å and d (Ge(2)-Ge(8)) =2.648 Å, significantly more than the bulk Ge distance of 2.432 Å. From the absence of dangling bond on the projected density of states plot in FIG. 17 (d) in the configuration with three-fold coordinated inner Ge atoms imply a double-bond between atoms 2 and 3 that then increases the distance to Ge (8). The distance between the inner and outer atoms are relatively more than that between inner atoms: d(Ge (1)-Ge (2))= 2.521 Å and d(Ge (3)-Ge (4))= 2.469 Å.

Since both the (3x1) and $\sqrt{3}$ models (discussed in sec. 5.6) have missing top layer reconstruction, they can transform from one to the other. In the optimized structure, the height of Ag from the average height of honeycomb-chain Ge atoms is 0.587 Å which is 0.727 Å in $\sqrt{3}$ structure.

## F. Ge(111)-Ag ($\sqrt{3} \times \sqrt{3}$) R30⁰

As a possible candidate of the $\sqrt{3}$ phase, we calculate the energetics of Ag on unreconstructed Ge(111) (defined above as IDL) and on the MTL reconstructed surface. We consider two models on reconstructed MTL, namely the honeycomb-chained-triangle (HCT) [2-4,95] of equilateral Ag triangles and the inequivalent-Ag triangle (IET) [5,6] that form a hexagonal network of two types of Ag triangles. The IET model with Ag triangles centered on T$_4$ sites is found to be energetically favored by 3.289 meV/Å$^2$ over the HCT (0.269 eV in energy which is close to the value of 0.20 eV [5]) and by 32.744 meV/Å$^2$ over Ag on IDL Ge(111) surface. The smaller size of Ag triangle in IET is attributed as the cause of the higher stability over HCT [5]. The top and side views of the IET model are shown in Fig. 18 (a) and (b), respectively.

Table I. Comparison of structural parameters (see FIG. 18) of the energetically favored Ge(111)-Ag ($\sqrt{3} \times \sqrt{3}$) R30⁰ IET model with the corresponding values from another theoretical study [5].

| Parameter | This study | Ref. [5] |
|---|---|---|
| $d_{Ag1}$ (Å) | 2.955 and 4.56 | 2.93 and 4.40 |
| $d_{Ag2}$ (Å) | 5.06 | 5.18 |
| $Z_{Ag}$ (Å) | 3.111 | 3.13 |
| $d_{Ge}$ (Å) | 2.685 | 2.82 |
| $\theta_{Ge}$ (degree) | 55.3 | 52 |

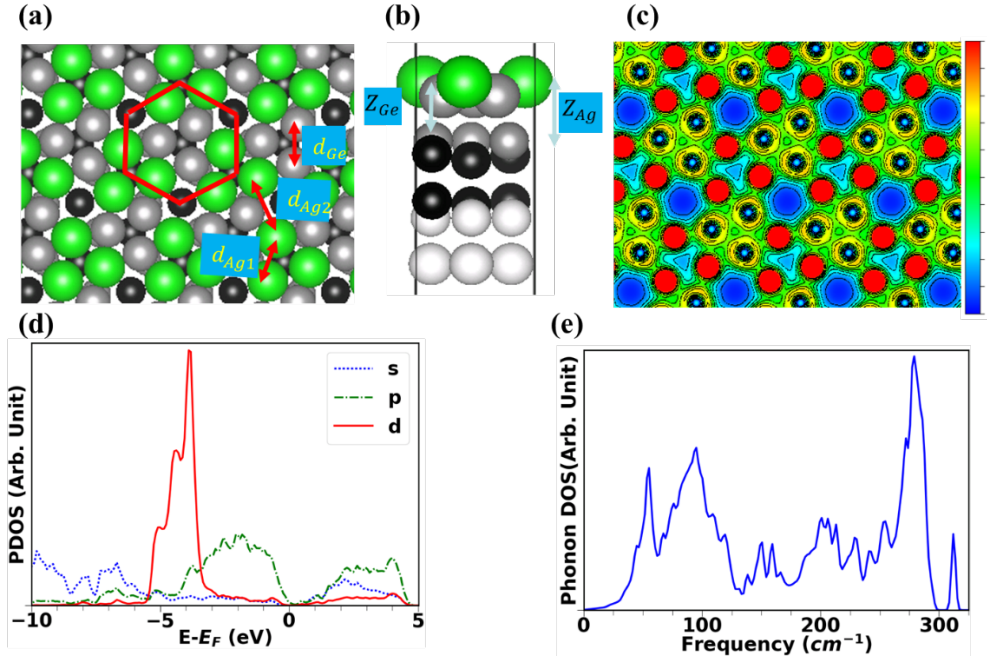

FIG. 18. (a) Top view of the repeated atomic model of the $\sqrt{3}$ phase showing a hexagonal structure formed by center of inequivalent triangles. (b) Side view. (c) Electron-density distribution in the (111) plane passing through the center of Ge atoms on MTL reconstructed layer on which center of light blue circles of small radius and red circles correspond to center of Ge atoms of MLR and projection of charge density at Ag atoms position on the plane, respectively (contours are plotted up to 0.08 $A_0^{-3}$ with the interval of 0.008 $A_0^{-3}$). (d) Projected density of electronic states. (e) Phonon density of states of the $\sqrt{3}$ phase. The green, gray and black spheres refer to Ag adatoms, and Ge atoms ($T_4$ and $H_3$) adsorption sites, respectively.

In the IET model, there are two equilateral triangular configurations of Ag centered about $T_4$ site whose side lengths ($d_{Ag1}$ in Fig. 18 (a)) are 2.955 Å and 4.56 Å, both being longer than the bulk Ag-Ag bond length of 2.932 Å. The distance between Ag adatoms that are in different triangles ($d_{Ag2}$ in FIG. 18 (a)) is 5.06 Å. The Ag layer is at height 0.727 Å from the MTL reconstructed Ge(111) surface, the value being close to the height difference between Ge layers on top Ge double layers of the IDL model. The top layer Ge atoms also form equilateral triangle centered about the $T_4$ site (see FIG. 18 (a)) with side lengths $d_{Ge}$ of 2.685 Å which is longer than bulk Ge bond length of 2.432 Å. Each Ge atoms of top MLR surface are at 2.473 Å from Ge atom at layer below, 2.684 Å from two Ge atom on the same layer and at 2.603 Å from two Ag atoms and 2.668 Å from third Ag adatom of Ag triangle. The interaction between these atoms can be seen on the charge density contour plot in FIG. 18 (c) taken on the (111) plane containing MLR Ge atoms. The Ge atoms are bonded with 2 of Ag atoms through open contours and charge distribution is higher in between Ge atoms indicating bonding between them. The coordination of Ge atoms of MLR with Ge and Ag atoms results into the absence of dangling bonds in the PDOS plot of the model in FIG. 18 (d).

The comparison of various structural parameters of the model from our calculation (see FIG. 18 (a) and (b) for definition) with those reported [5] is presented in Table I, which show close agreement. The phonon DOS plot shown in FIG. 18 (e) imply the thermodynamic stability of the structure.

## G. Ge(111)-Ag (4×4) phase

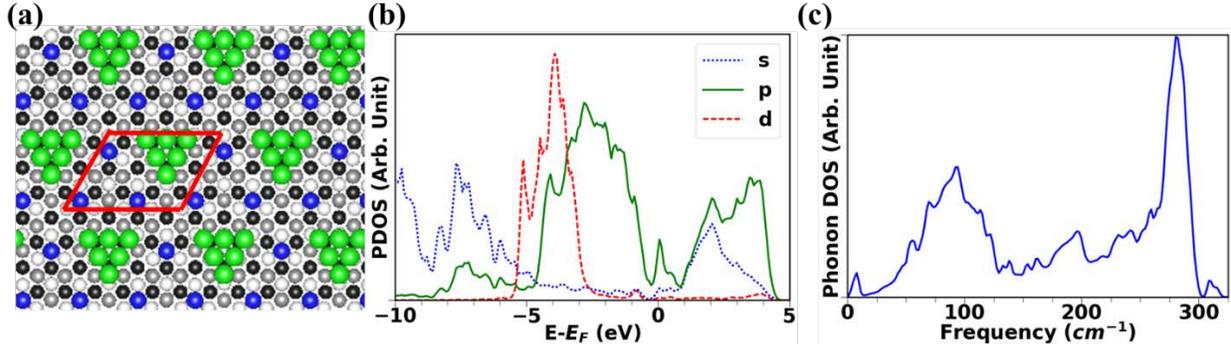

FIG. 19. (a) Top view of the atomic model of the Ge(111)-Ag (4×4) phase with its surface unit cell represented by parallelogram where the green, blue, white, gray, and black spheres refer to Ag adatom, Ge adatom, Ge atom which are the $T_1$, $T_4$, and $H_3$ adsorption sites, respectively. (b) Orbital projected electronic density of states. (c) Phonon density of states.

Of the proposed models for the (4×4) phase by Hammar et al. [2], Spence et al. [7], Weitering et al. [9] and Collazo et al. [8], we find that by Hammar et al. [2] to be the most energetically favored (see FIG. 19 (a) for top view). The model contains six Ag adatoms in one triangular subunit and three Ge adatoms on another triangular subunit adsorbed on IDL Ge(111) surface. The heights of Ag adatoms on corners and edges from Ge atoms on layer below them are 2.148 Å and 2.456 Å, respectively. The corner and edge Ag adatoms form equilateral triangles with length of 5.639 Å and 2.844 Å, respectively. The distance from Ag on corner to the nearest neighbor Ag on edge is 2.822 Å and that between Ag on edges is 2.846 Å, both distances are less than their distance in bulk implying that the Ag adatoms form bonded structure. The Ag adatoms are not at top site as reported [2], rather at an angle of $30.6^0$ and $15.5^0$ from the normal drawn on Ge atom on layer below. The three Ge adatoms bond with nine of the remaining ten surface Ge atoms keeping one as rest atom which pops up the surface as in Ge(111)–c(2×8) by height 0.618 Å. As in the Ge(111)–c(2×8) model, the surface Ge atoms just below Ge adatoms are pushed into the bulk by height 0.591 Å as compared to other Ge atoms on the same layer. The presence of Ge adatoms and rest atoms contribute to dangling bond at Fermi level, which one can see in the orbital resolved density of electronic states in FIG. 19 (b).

The thermodynamic stability of the phase is shown using the phonon density of states plot in FIG. 19 (c). The small feature at about 6 cm$^{-1}$ corresponds to the back and forth vibration of the Ag cluster about its geometric center parallel to the surface. This low frequency in-plane vibration is weakly coupled to the substrate, as in the case of Co clusters on Cu(111) [96]. The feature at high frequency (320 cm$^{-1}$) corresponds to bond stretching vibrations of Ge atoms below the Ge adatoms. Although there is no experimental report of the vibrational study of the phase, the appearance of such features in the calculated phonon density of states attest to the presence of Ag clusters and Ge adatoms on the T$_4$ site.

As explained in computational details, the phonon density of states of each of the stable phases are used in equation (2) to calculate $F_{slab}^{vib}(T)$ as part of $\gamma(T)$ in equation (1), which is used to construct surface phase diagram as explained below in section 5.8.

## H. Ag/Ge(111) surface phase diagram

Up to this point, we have shown that Ge(111)-c(2×8), Ge(111)-Ag (3×1), Ge(111)–Ag ($\sqrt{3} \times \sqrt{3}$) R30$^0$ and Ge(111)-Ag(4×4) are the energetically favored and thermodynamically stable phases. We next construct the surface phase diagram by considering stable isolated phases and the possible multi-phase coexistence, by evaluating the surface free energy of a combination of phases for given coverage (u) and temperature (T) using

$$\gamma(T, u) = a_0 \gamma_{Ge-c(2\times8)}(T) + a_1 \gamma_{3\times1}(T, u) + a_2 \gamma_{\sqrt{3}}(T, u) + a_3 \gamma_{4\times4}(T, u) \tag{4}$$

where $a_0, a_1, a_2, a_3$ refer to the value of coverage of phases: $a_0$ of Ge c(2×8), $a_1$ of Ag-(3 × 1), $a_2$ Ag-($\sqrt{3} \times \sqrt{3}$) R30$^0$ and $a_3$ of Ag-(4 ×4). They are required to satisfy the condition that the total fraction of possible phases sum to unity i.e.,

$$a_0 + a_1 + a_2 + a_3 = 1, \tag{5}$$

and the given Ag coverage (u) can be expressed as combination of possible Ag phases as

$$u = \frac{1}{3}a_1 + a_2 + + \frac{6}{16}a_3 \tag{6}$$

where the coefficients $\frac{1}{3}$, 1 and $\frac{6}{16}$ of $a_1$, $a_2$ and $a_3$, respectively refers to Ag coverage on surface unit on $\sqrt{3}$, (3 × 1) and (4 × 4) phases, respectively. For each coverage and temperature, possible combinations are considered, and the one that minimizes the value of γ (T, u) in equation (4) and satisfies equations (5) and (6) is taken as the calculated phase for that set of conditions. The coverage-temperature dependent surface phase diagram of Ag/Ge(111) system obtained using the outlined approach is shown in FIG. 20.

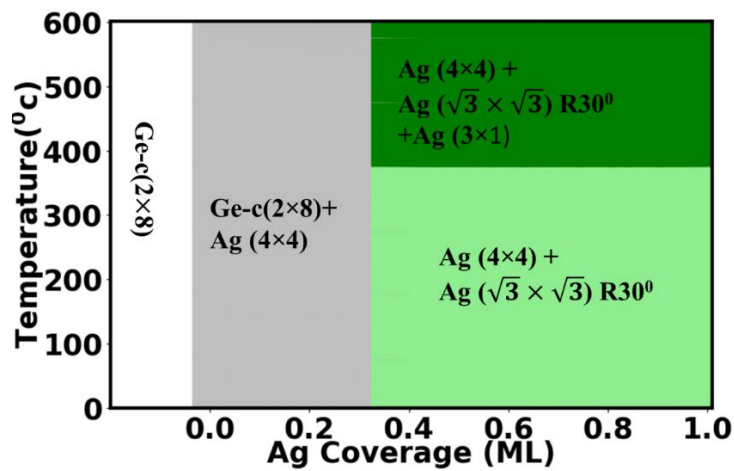

FIG. 20. Coverage-temperature dependent surface phase diagram of Ag/Ge(111).

The Ge(111)-c(2×8) phase is found to have the lowest surface free energy at all temperatures under study over the dynamically stable reconstructed Ge(111)-2×1 phase. The (4× 4) overlayer Ag structure coexisting with Ge(111)-c(2×8) has the minimum surface free energy until its saturation coverage of 0.375 ML. At temperature below 400°C and at coverage >0.375 ML, the (4×4) structure coexists with $\sqrt{3}$ structure because the combination has the lowest surface energy, and the proportion of the latter structure keeps increasing with increasing Ag coverage until 1 ML. At Ag coverage beyond 0.375 ML and temperature beyond 400°C, with increasing Ag coverage, the proportion of (4×4) decreases following the same trend as below 400°C and the combined proportion of $\sqrt{3}$ and (3×1) increases with a small presence of the (3×1) phase with maximum to 2% of total coverage.

The above conclusion that the c(2×8) phase is the energetically favored one is in agreement with the experimental results shown in the phase diagram in FIG. 7. Although the computational phase diagram does not include Ge(111)-(2×1) because of its slightly higher surface free energy over the c(2×8) (see FIG. 15), the small difference in their surface free energy agrees with the experimental observation of 2×1 coexisting above 300$^0$C. Note that our computational results find both of those phases as dynamically stable. The coexistence of (4×4) and the Ge(111)-c(2×8) phase up to 0.375 ML is in agreement with the experimental phase diagram. The large region of the phase diagram showing the coexistence of (4×4) with the $\sqrt{3}$ phase for coverage >0.375 ML agrees with the experimental phase diagram, in which with increasing Ag coverage, the fraction of the $\sqrt{3}$ phase increases until its saturation coverage of 1 ML. The (3×1) phase is calculated to coexist with other phases only at coverage beyond 0.375 ML; this result differs from the experimental observation of the (3x1) phase coexisting for lower coverages. Consistent with the experimental observation of the (1×1) phase around 600$^0$C in disordered phase, our computational study find it to be dynamically unstable.

## VI. CONCLUSIONS

We show the experimentally determined structures of Ag adatoms on the Ge(111) surface using low energy electron diffraction, low energy electron microscopy, and scanning tunneling microscopy as functions of coverages and temperature. The atomic models of those phases are determined using the density functional theory approach. For each of structures, the electronic structures are presented to find the physical reasons for their formations, and the vibrational structures are calculated to determine their thermodynamical stabilities.

In the energetically favored intact double layer model of the unreconstructed Ge(111) surface, the top layer atom has a dangling bond of $sp_z$ type and the surface is dynamically unstable due to acoustic modes of vibration. The intact double layer Ge(111) surface with four Ge adatoms adsorbed on the three-fold hollow $T_4$ site in each surface unit cell is found to be the model for the reconstructed Ge(111) surface. The inequivalent trimer model with Ag adatoms on the missing top layer reconstructed Ge(111) surface at 1 ML coverage is the favored atomic model for the $\sqrt{3}$ phase. The honeycomb-chain channel model of the (3×1) phase in which the missing top layer reconstructed Ge atoms form a honeycomb with metal atoms in the channel is found to be the atomic model of the (3×1) phase. The atomic model of the (4×4) phase is metallic with two triangular subunits with six Ag adatoms on one subunit and three Ge adatoms on $T_4$ sites on another subunit at different height. Our vibrational study confirms that the atomic models corresponding to the experimentally observed phases are dynamically stable.

An experimental surface phase diagram is produced as a summary of the detailed observations of the multiple surface structural phases as a function of temperature and coverage, and a similar diagram is constructed by minimizing the surface free energy of the combination of stable phases. In agreement with experimental observations, the computational study show Ge(111)-c(2×8) as the favored Ge(111) phase at low temperature. The coexistence of the (4×4) phase with Ge(111)-c(2×8) phase up coverage of 0.375 ML agrees in both the experimental and theoretical phase diagrams, as does the coexistent of the (4×4) phase with the $\sqrt{3}$ phase at coverage >0.375 ML at a range of temperatures. The calculations show coexistence of (3×1), $\sqrt{3}$, and (4x4) at temperatures beyond 400$^0$C, while the experiment shows coexistence of (3x1) and $\sqrt{3}$ above 500$^0$C. Both experiment and theory show the (1×1) phase as a disordered structure formed only at the desorption temperature.

## ACKNOWLEDGMENTS

We would like to acknowledge the Extreme Science and Engineering Discovery Environment (XSEDE) through allocation DMR130009 and the Advanced Computing Center at the University of Central Florida for high-performance computational resources. We are pleased to acknowledge support from the National Science Foundation under Grants DMR-1710306 (S.R.A., D.L. and T.S.R.) and DMR-1710748 (S.C.). We thank Professor C. Y. Fong for critical reading of the manuscript.